\documentclass{jfm}
\usepackage{graphicx}
\usepackage{epstopdf, epsfig}
\usepackage{latexsym,bm}
\usepackage{longtable}

\usepackage{xcolor}

\usepackage{graphicx}
\usepackage{dcolumn}
\usepackage{bm}
\usepackage{hyperref}
\hypersetup{
    colorlinks=true,
    linkcolor=blue,
    filecolor=blue,      
    urlcolor=blue,
    citecolor=blue,
}

\shorttitle{Triadic resonances in rotating convection}
\shortauthor{Y. Lin}
\title{Triadic resonances driven by thermal convection in a rotating sphere }

\author{Yufeng Lin\aff{1}
	\corresp{\email{linyf@sustech.edu.cn}}}

\affiliation{\aff{1} Department of Earth and Space Sciences, Southern University of Science and Technology, Shenzhen 518055, China}

\begin{document}
\maketitle
\begin{abstract}
 We report for the first time on triadic resonances in a rotating convection system. Using direct numerical simulations, we find that convective modes in a rotating spherical fluid can excite a pair of inertial modes whose frequencies and wavenumbers match the triadic resonance conditions. Depending on the structures of the convective modes, triadic resonances can lead to the growth of either a pair of modes with lower frequencies and wavenumbers, or a pair of modes with higher frequencies and wavenumbers, providing a possible mechanism for the bi-directional energy cascade. Increased thermal forcing leads to fully developed turbulence, which also exhibits wave-like motions, and is reminiscent of the energy spectrum of inertial wave turbulence. Our results suggest that the interaction of inertial waves plays an important role in rotating convection, which is of great importance in understanding the dynamics of planetary and stellar interiors.
\end{abstract}
\section{Introduction}\label{sec:Introduction}
Fluid flows in stars and planets, which are typically subject to rotation, can be driven by internal thermal forcing 
\citep{Spiegel1971,Aurnou2015,Jones2015} and/or by external mechanical forcings due to the gravitational coupling with orbital companions \citep{LeBars2014,Ogilvie2014}. Despite the common rotational constraint, thermal convection and mechanically driven flows are traditionally  treated as two separate branches in rotating fluid dynamics \citep{Chandra1961book,Greenspan1968book}.  
 
Owing to the periodicity of orbital motions, mechanical forcings such as precession \citep{Hollerbach1995,Noir2001b}, libration \citep{Aldridge1969} and tides \citep{Ogilvie2004} can drive oscillatory motions in rotating fluids, known as inertial modes or inertial waves restored by the Coriolis force \citep{Greenspan1968book}. On the other hand, it has been long recognized that rotating convection close to the onset is in the form of quasi-steady columnar rolls \citep{Busse1970,Jones2000}. At low Prandtl number ($\Pran < 1$), however,  it was found that the onset of convection can be in the form of an oscillatory inertial mode at leading order \citep{Zhang1994}, suggesting that rotating convection and mechanically driven rotating flows are closely related. Indeed, a recent monograph by  \cite{Zhang2017} provided a set of unified asymptotic theories for the linear onset of convection and linear responses to mechanical forcings in rotating fluids based on an expansion of inertial waves. In this paper, we show that some aspects of the nonlinear dynamics of rotating convection can be understood within the framework of inertial wave interactions, as developed in mechanically driven rotating flows \citep{Kerswell2002,LeBars2014}.

Rotating flows can become unstable through the so-called triadic resonance \citep{LeBars2014}, i.e. a base flow with frequency $\omega_0$ and azimuthal wavenumber $m_0$ can excite two inertial waves whose frequencies $\omega_1,\, \omega_2$ and azimuthal wavenumbers $m_1,\,m_2$ match the resonance conditions { \citep{Bretherton1964,Kerswell2002}:} (Here we present the resonance conditions in a spherical geometry. In an unbounded domain, the second condition should read $\bm{k}_1\pm \bm{k}_2=\bm{k}_0$, where $\bm k$ is the wavevector.)
\begin{equation}
\omega_2\pm \omega_1=\omega_0, \quad m_2\pm m_1=m_0.
\end{equation}
 Such an instability has been observed both experimentally and numerically in a variety of rotating flows driven by mechanical forcings \cite[see][and references therein]{Kerswell2002,LeBars2014}, but never reported before in rotating convection systems. Motivated by geophysical and astrophysical applications, rotating convection has been studied extensively both for planar models \cite[see recent review by][]{Plumley2019} and for spherical models  \cite[e.g.][]{Aubert2001,Christensen2002,Gastine2016,Kaplan2017,Guervilly2019}, 
yet the nonlinear dynamics and turbulent regime remain to be fully understood \citep{Aurnou2015}. Recent numerical simulations \citep{Horn2017} and  laboratory experiments \citep{Aurnou2018} of rotating convection in a cylinder using liquid gallium ($\Pran \approx 0.025$) observed multi-modal interactions, yet the underlying mechanism remains to be elucidated. { \cite{Lam2018} numerically studied rotating convection in a sphere with $Pr=0.0023$ at moderate Ekman numbers and found that the nonlinear interaction of various thermal inertial waves leads to a weakly turbulent state, though no evidence of triadic resonances was shown.} Here we show, through numerical simulations { at lower Ekman numbers}, that the primary convective mode at low $\Pran$ (relevant to liquid metals) can excite a pair of inertial modes through the triadic resonance in a rotating sphere, leading to more complex convective motions. Increasing the thermal forcing leads to a turbulent state that shows signatures of inertial waves. Our study may offer a new pathway to study rotating turbulent convection within the framework of inertial wave turbulence \citep{Godeferd2015,LeReun2017}.   

\section{Numerical model}

We consider the canonical problem of Boussinesq convection in a sphere of radius $r_o$, which rotates at $\bm \Omega=\Omega \bm{\hat z}$. Convection is driven by a uniform heat source $S$ under a gravitational field $\bm g=-g_o\bm r/r_o$. The problem was first formulated by \cite{Chandra1961book}  and is governed by the following dimensionless equations in the rotating frame:
\begin{equation} \label{eq:NS}
E \left(\frac{ \partial \bm u}{\partial t}+\bm {u\cdot \nabla u}\right)+\bm{\hat z}\times \bm u=-\bm \nabla p +\frac{ERa}{Pr}T\bm r+E\nabla^2 \bm u,
\end{equation}
\begin{equation}\label{eq:Temp}
Pr \left(\frac{ \partial T}{\partial t}+\bm{u\cdot \nabla} T \right)=\nabla^2 (T-T_b),
\end{equation}
\begin{equation} \label{eq:divu}
\bm{\nabla \cdot u}=0,
\end{equation}
where $\bm u$ is the fluid velocity, $p$ is the reduced pressure, $T$ is the total temperature and $T_b=(1-r^2)/2$ is the dimensionless basic temperature without convection. The system is defined by three non-dimensional parameters, i.e. the Ekman number $E$, the Rayleigh number ${Ra}$ and the Prandtl number $\Pran$:
\begin{equation}
E=\frac{\nu}{2\Omega r_o^2}, \quad Ra=\frac{\alpha g_o \beta  r_o^5}{\nu \kappa}, \quad \Pran=\frac{\nu}{\kappa},
\end{equation}
where $\nu$ is the kinematic viscosity, $\kappa$ is the thermal diffusivity, $\alpha$ is the thermal expansion coefficient and { $\beta=S/(3\kappa)$ }characterizes the temperature gradient without convection.  { In rotating flows, another important non-dimensional parameter is the Rossby number, which is defined as
\begin{equation}
Ro=\frac{U_{rms}}{2\Omega r_o}.
\end{equation}
In our model, the Rossby number $Ro$ is an output parameter and is determined by the dimensional root mean square velocity $U_{rms}$ in the simulation domain.
}
 
The fully nonlinear governing equations of the system are numerically solved using a spectral method based on expansions of spherical harmonics in the angular direction and Jones-Worland polynomials in the radial direction \citep{Marti2016}. We use the no-slip boundary condition and a fixed-temperature ($T=0$) condition on the spherical boundary. Simulations are initiated from small random velocity perturbations or from a saturated state at lower $Ra$. 

This study explores rotating convection in the parameter regime of low Ekman numbers ($10^{-6}\le E\le 10^{-5}$) and low Prandtl numbers ($10^{-3}\le Pr\le 10^{-2}$). The Rayleigh number varies from the slightly supercritical to the turbulent convection regime. The resulting Rossby number ranges from $10^{-3}  $ to $3\times 10^{-2}$, suggesting that convection is rotationally dominated in all of the simulations. We use numerical truncations of $N=63$, $L=M=127$ for the simulations near the onset, and increase the truncations up to $N=191$, $L=M=383$ for the turbulent convection, where $N$, $L$ and $M$ are truncations of the radial polynomials, degree and order of spherical harmonics, respectively. The convergence of the numerical solutions is warranted by checking energy spectra and by doubling the truncations for some cases.

\section{Results}
\begin{figure}
\begin{center}
\includegraphics[width=0.8\textwidth]{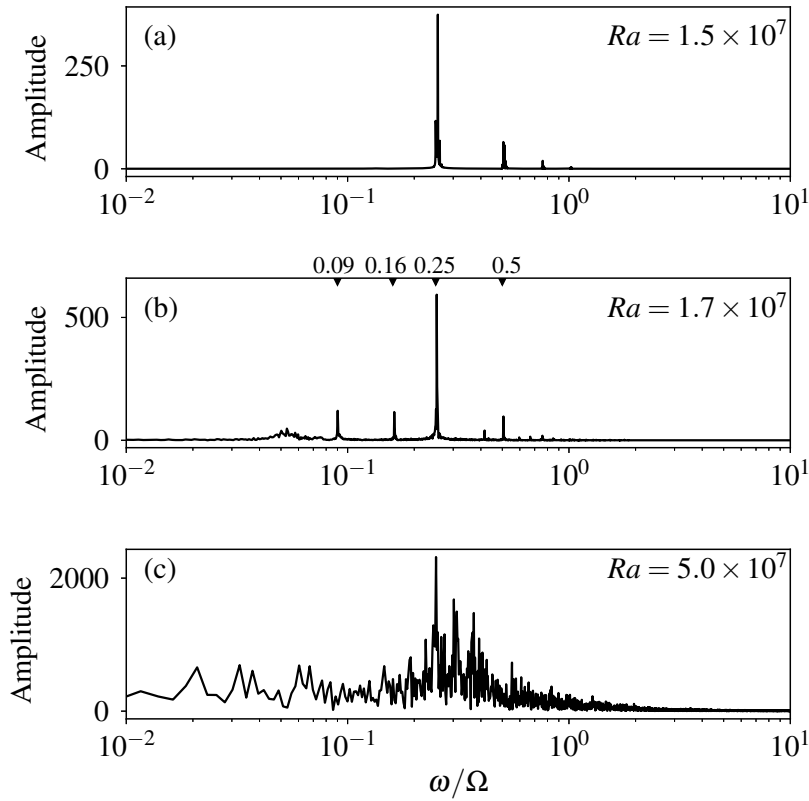}
\end{center}
\caption{Fourier transform of the radial velocity $u_r$ at a fixed position ($r=0.59$) in the equatorial plane at $E=10^{-6}$ and $\Pran=0.001$, but various $Ra$. Frequencies are normalized by the rotation frequency $\Omega$.}
\label{fig:VelFFT}
\end{figure}

Figure \ref{fig:VelFFT} shows the Fourier transform of the time series of the radial velocity at a fixed position ($r=0.59$) in the equatorial plane for various $Ra$ at $E=10^{-6}$ and $\Pran=0.001$. For $Ra=1.5\times 10^{7}\approx1.4 Ra_c$, where $Ra_c$ is the critical Rayleigh number for the onset of convection, the frequency spectrum shows a pronounced peak at $\omega/\Omega=0.25$, which corresponds to an oscillatory convective mode at low $\Pran$  \citep{Zhang1994}. In figure \ref{fig:VelFFT} (a), there is also an obvious peak  at $\omega/\Omega\approx0.5$, twice the frequency of the primary mode. Similar doubling in frequency has been observed  in recent numerical simulations \citep{Horn2017}  and laboratory experiments \citep{Aurnou2018}  using liquid metal in a rotating cylinder . As we shall show later, the doubling in frequency is accompanied by a doubling in the azimuthal wavenumber as well, which may result from the nonlinear self-interaction of the primary mode. (The nonlinear term of a flow $\bm u \propto \mathrm e^{\mathrm i (\omega t+m \phi)}$ potentially has two components: $m_0=0,\,\omega_0=0$ and $m_2=2m,\,\omega_2=2\omega$). For $Ra=1.7\times 10^{7}\approx1.6Ra_c$,  apart from similar peaks observed in figure \ref{fig:VelFFT}(a), the spectrum exhibits two distinct peaks with similar amplitudes at $\omega/\Omega=0.09$ and 0.16, respectively  ({ figure} \ref{fig:VelFFT}(b)). These two peaks and the primary peak at $\omega/\Omega=0.25$ satisfy the triadic resonance condition, i.e.  $\omega_1 \pm \omega_2=\omega_0$, suggesting a possible resonant triad between the primary convective mode and a pair of free modes.

\begin{figure}
\includegraphics[height=0.32 \textwidth]{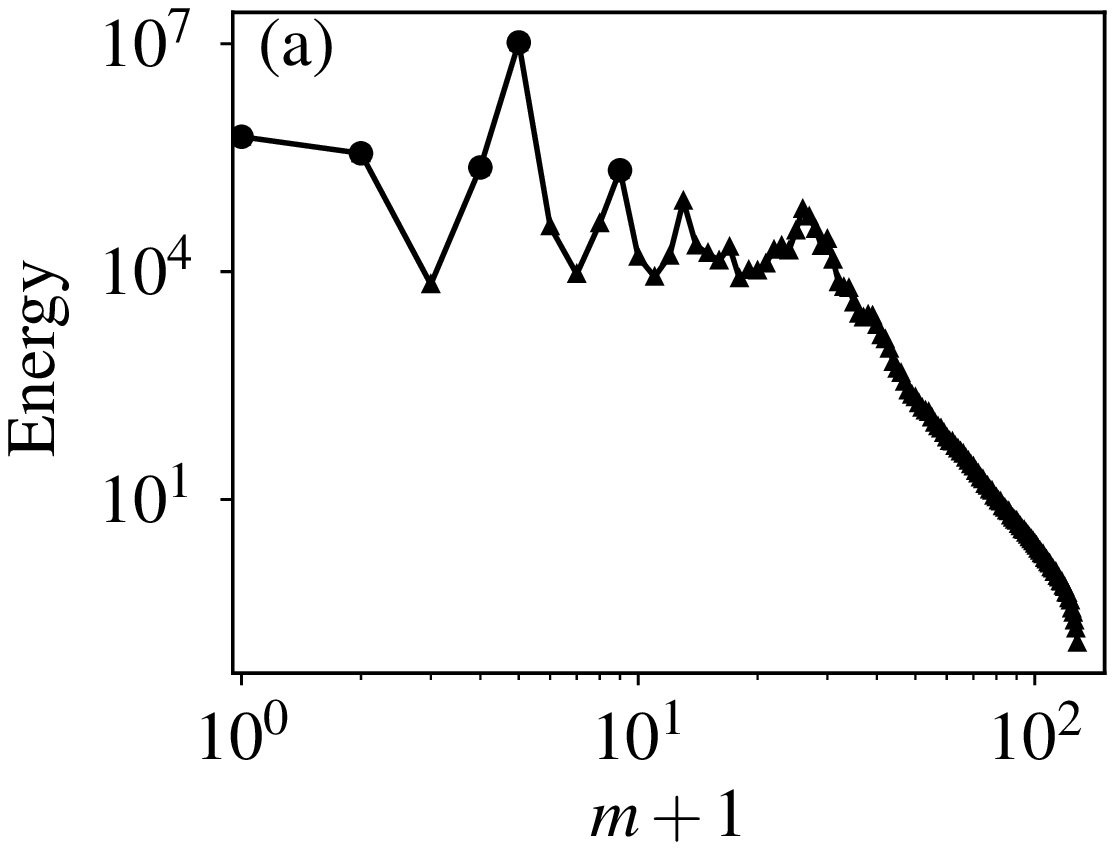}	
\includegraphics[height=0.32 \textwidth]{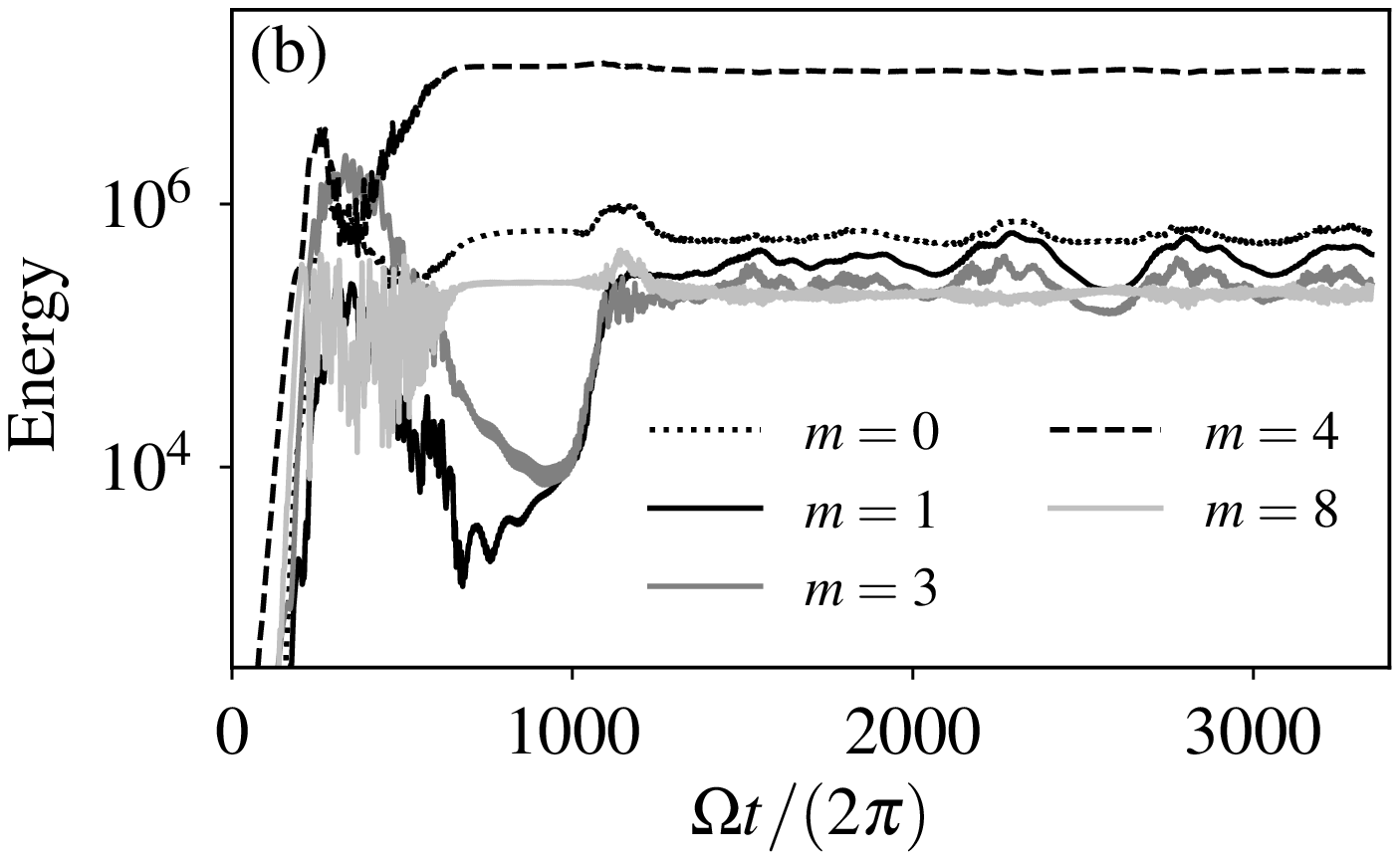}\\
\caption{(a) Time-averaged kinetic energy spectrum as a function of the azimuthal wavenumber $m$ at $E=10^{-6}$, $\Pran=0.001$ and $Ra=1.7\times10^7$. (b) Time evolution of the kinetic energy contained in different $m$, for the five largest components { (represented by circles)} in panel (a). Time is in the unit of the rotation period.}
\label{fig:KEm} 
\end{figure}

Figure \ref{fig:KEm} shows the kinetic energy contained in different azimuthal wavenumbers $m$ for the case in figure \ref{fig:VelFFT}(b). We can see from the time-averaged spectrum that the energy is dominated by the $m=4$ component, with subsequent major { contributions} from the $m=0,1,3$ and 8 components. Figure \ref{fig:KEm}(b) shows the time evolution of the kinetic energy for the aforementioned components. After a transient stage, the $m=4$ component starts to grow and becomes the dominant one, which corresponds to the primary convective mode. 
As the primary mode becomes saturated, the $m=0$ and $m=8$ components also develop and { saturate}. The $m=0$ component represents a steady zonal flow, while $m=8$ is twice the azimuthal wavenumber of the primary mode. Both $m=0$ and $m=8$ components can be attributed to the  nonlinear interaction of the convective mode with itself \citep{Zhang2017}. 
More interestingly, at { $\Omega t/(2\pi)\approx 1000$}, the $m=1$ and $m=3$ components start to grow exponentially with the same growth rate and then saturate to similar levels, again  pointing to a possible triadic resonance between the primary mode $m=4$ and two free modes of $m=1$ and $m=3$. 

\begin{figure}
\begin{center}
	\includegraphics[width=0.7 \textwidth]{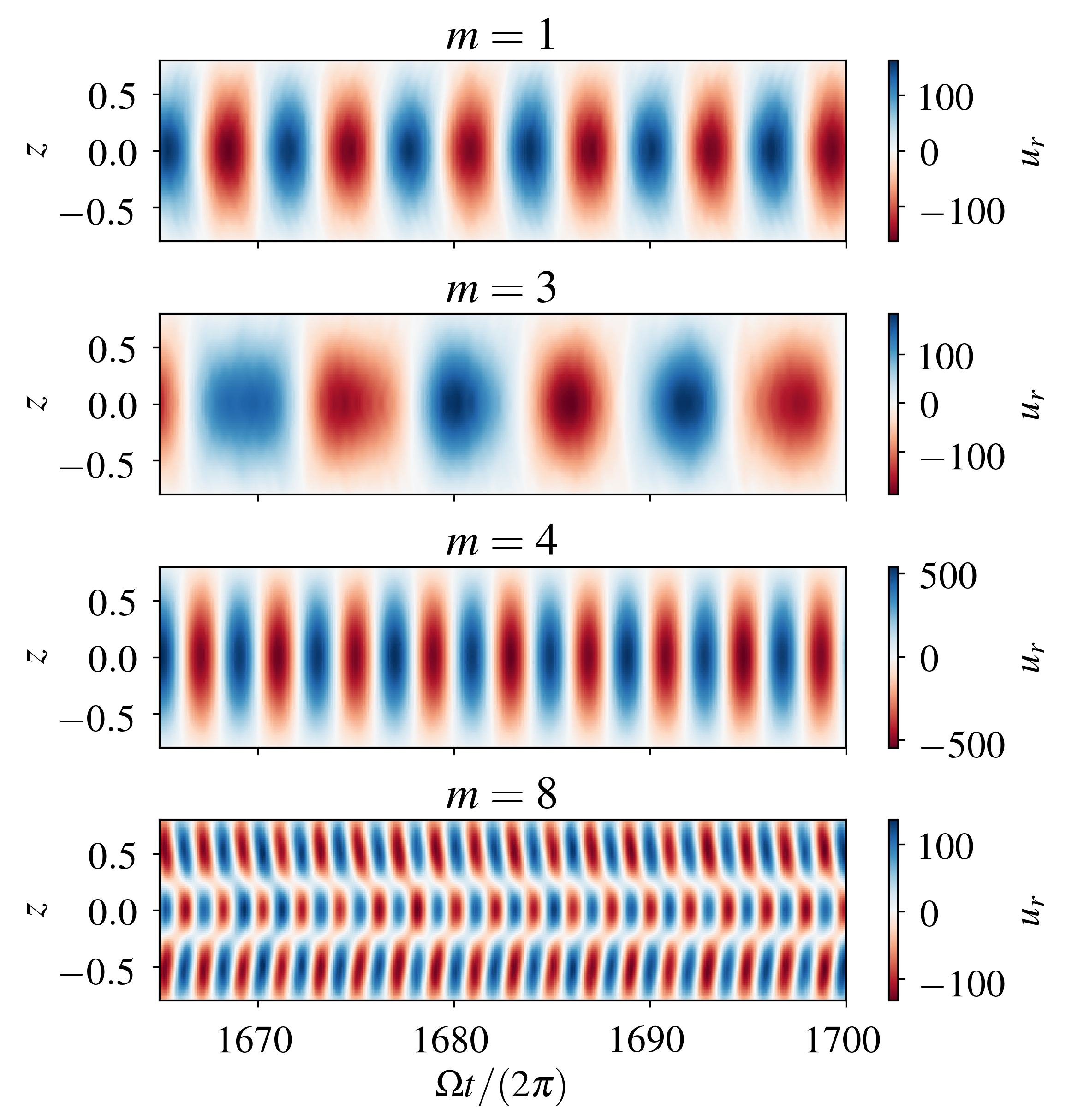}
	\end{center}
	\caption{Time evolution of the radial velocity  along a fixed vertical line with the distance to the rotation axis $s=0.5$ for the case in Fig. \ref{fig:KEm}. The velocity field has been decomposed into components with different azimuthal wavenumber $m$.}
	\label{fig:VelLine}
\end{figure}

In order to confirm that both $\omega$ and $m$ simultaneously  satisfy the triadic resonance conditions, we decompose the velocity field into different $m$ components and then analyse the time evolution of each component.
Figure \ref{fig:VelLine} shows the time evolution of the radial velocity along a fixed vertical line for $m=1,3,4$ and 8 components, respectively, for the case in figure \ref{fig:KEm}. { As the $m=0$ component represents a steady zonal flow, it is not shown in figure \ref{fig:VelLine}.} One can see that each $m$-component corresponds to a single frequency mode, and the $m=1,$ 3, 4 modes have the simplest structure along the axis of rotation. We extract the frequencies of each { mode} by fitting the time series with a sinusoidal function and find  $\omega/\Omega=0.1624,$ 0.0899, 0.2529 for the $m=1$, 3, 4 modes respectively (see Table \ref{tab:Triads}), which clearly satisfy the triadic resonance conditions in both $\omega$ and $m$. These results convincingly demonstrate that the primary convective mode with $m=4$ excites two free modes with $m=1$ and 3 { through} the triadic resonance. We also find that the $m=8$ mode has a frequency of $\omega/\Omega=0.5058$, doubling both $m$ and $\omega$ of the primary mode, as we mentioned before.

\begin{table}
\begin{center}
\def~{\hphantom{0}}
\begin{tabular}{ccccrrrrrrrrr}
			$E$ & $\Pran$& $Ra$&Ro & \multicolumn{3}{c}{Primary mode} &  \multicolumn{3}{c}{Free mode 1}  & \multicolumn{3}{c}{Free mode 2}  \\		
			($\times 10^{-6}$)& & ($\times 10^{6}$)& ($\times 10^{-3}$) &$m$ & $\omega_{\mathrm{ns}}$ & $\omega_{\mathrm{in}}$& $m$ & $\omega_{\mathrm{ns}}$  & $\omega_{\mathrm{in}}$& $m$ & $\omega_{\mathrm{ns}}$ & $\omega_{\mathrm{in}}$ \\
			\vspace*{0.01cm} \\
		1 &0.001 & 17  & 2.4 & 4 & 0.2529 & 0.2613 & 1 & 0.1624 & 0.1766 & 3 & 0.0899 & 0.0715\\ 
		5 &0.001 & 1.9 &9.6& 1 & 0.1781 & 0.1766 & 7 & -0.5352 & -0.5380 & 8 & -0.3565 & -0.4196\\
			10 &0.01 & 3.3 &2.2& 3 & 0.0918 & 0.0715 & 10 & -0.3596 & -0.3771 & 13 & -0.2678 & -0.3303 \\
		\end{tabular}
\caption{\label{tab:Triads}
Examples of the triadic resonance at different control parameters and involved modes. $\omega_{\mathrm{ns}}$ is the frequency obtained from numerical simulations and $\omega_{\mathrm{in}}$ is the frequency of purely inviscid inertial modes. Frequencies are in the unit of the rotation frequency and positive (negative) values correspond to prograde (retrograde) modes. For each case, a movie showing the radial velocity of the interacting modes in the equatorical plane is also provided.}
\end{center}
\end{table}

\begin{figure}
\begin{center}
	\includegraphics[width=0.7 \textwidth]{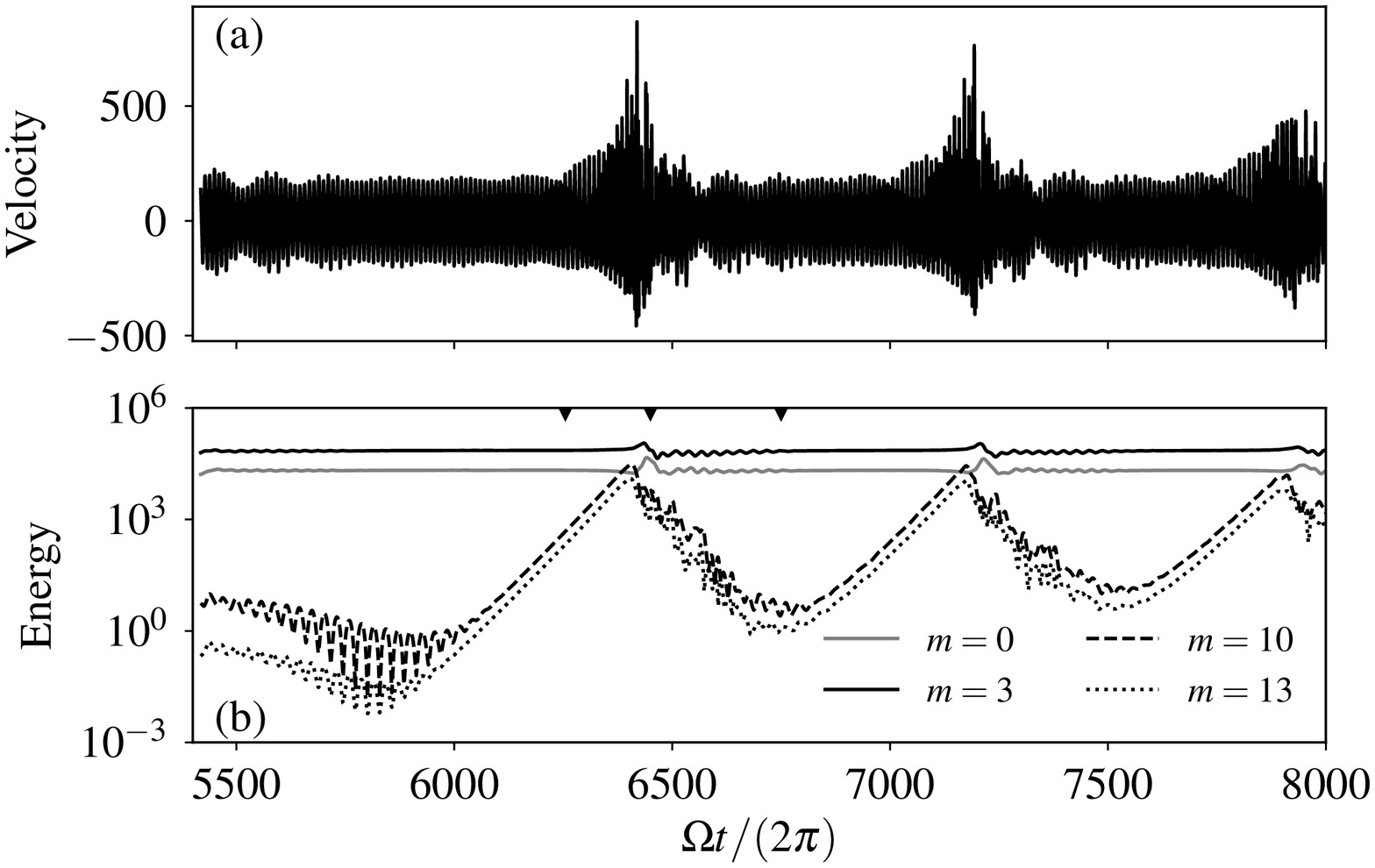} \\
	\includegraphics[width=0.32 \textwidth,clip,trim=0 0 8cm 0]{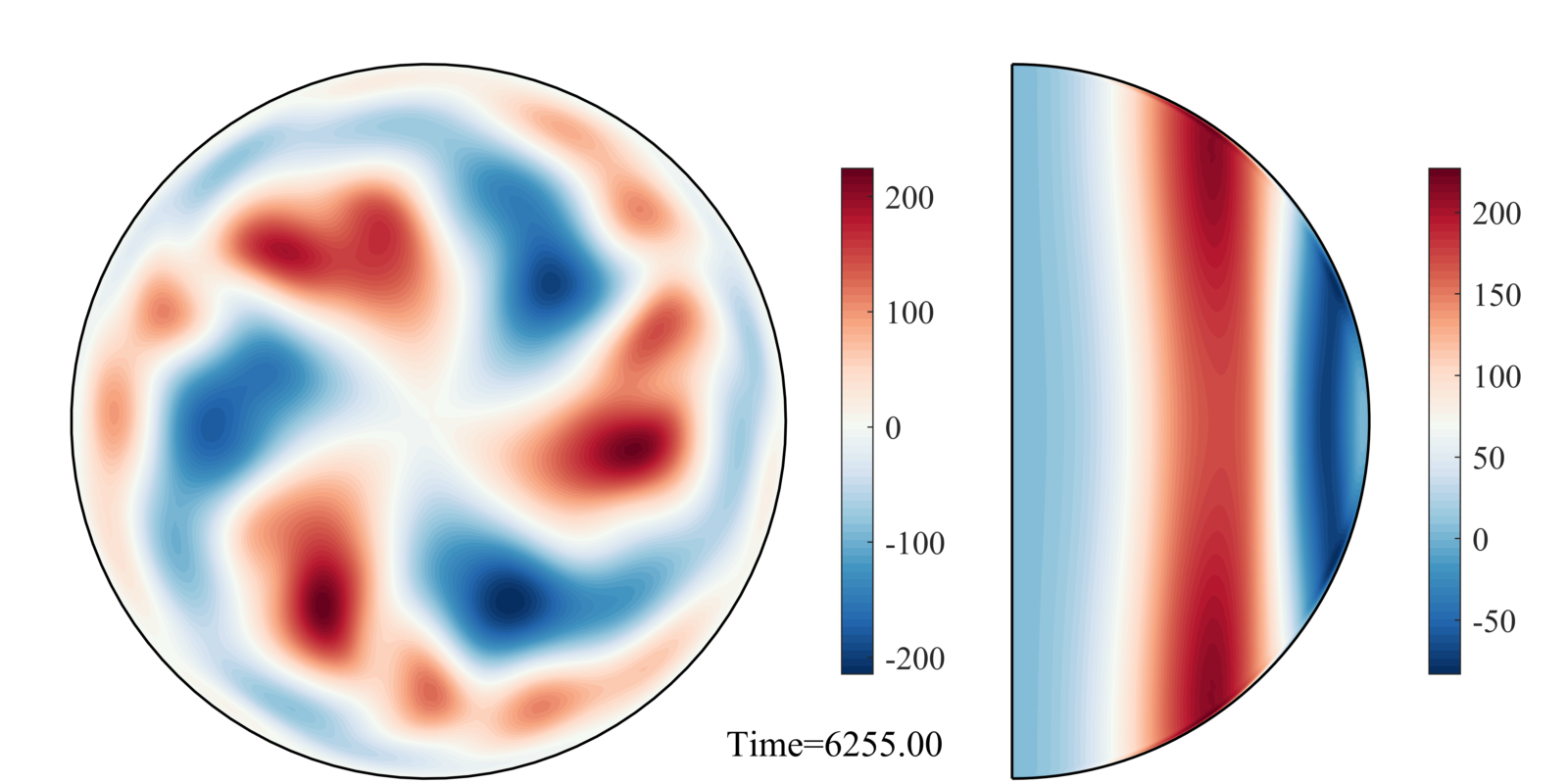} 
	\includegraphics[width=0.32 \textwidth,clip,trim=0 0 8cm 0]{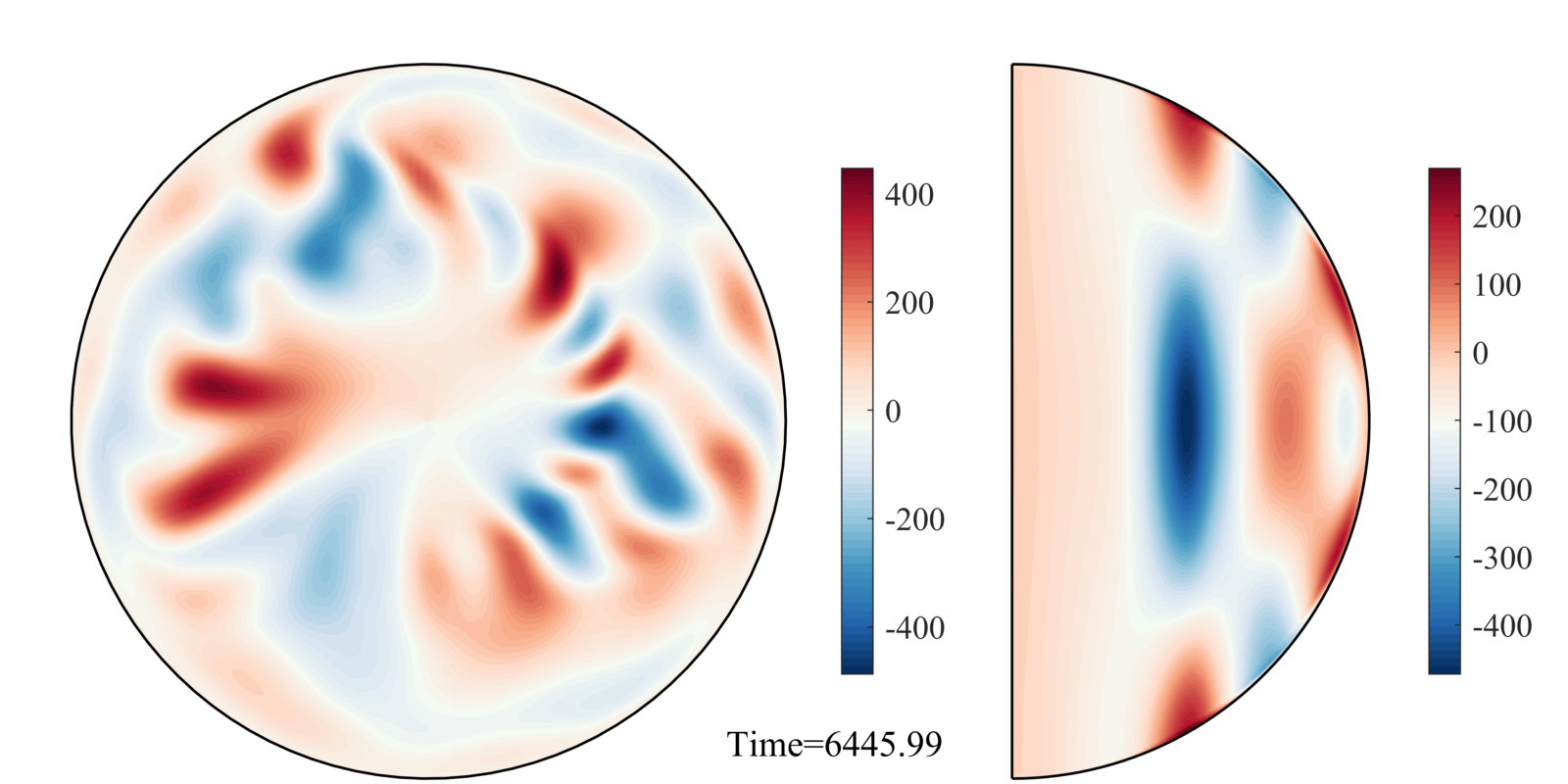}
	\includegraphics[width=0.32 \textwidth,clip,trim=0 0 8cm 0]{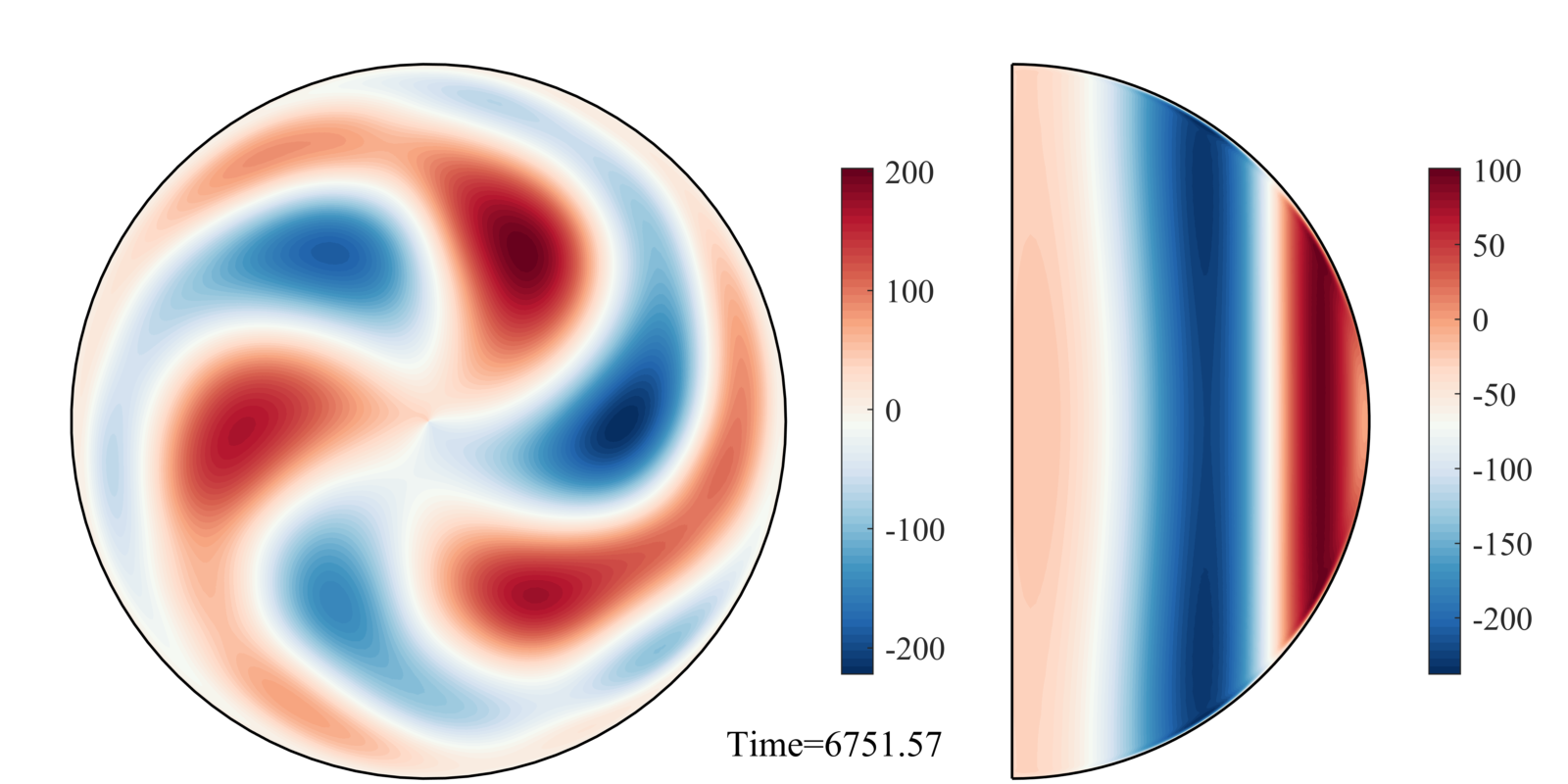}\\
	(c) \hspace*{4cm}(d) \hspace*{3.8cm}(e) 
	\end{center}
	\caption{The growth and collapse of the unstable modes at $E=10^{-5}$, $\Pran=0.01$ and $Ra=3.3\times 10^6$. (a)Time series of the radial velocity at a fixed point. (b) Time series of the kinetic energy contained in different $m$ components; (c-e) Snapshots of the radial velocity in the equatorial plane at instances indicated by the inverted black triangles in panel (b).}
\label{fig:GrowthCollapse}
\end{figure}

Table \ref{tab:Triads} lists three examples of triadic resonances and involved modes (see also supplementary \href{http://faculty.sustech.edu.cn/wp-content/uploads/2020/11/2020111814352427.mp4}{Movie 1}, \href{http://faculty.sustech.edu.cn/wp-content/uploads/2020/11/2020111814352598.mp4}{Movie 2}, \href{http://faculty.sustech.edu.cn/wp-content/uploads/2020/11/2020111814352655.mp4}{Movie 3} for the radial velocity of interacting modes in the equatorial plane) at different control parameters.   We can see that the triadic resonance can excite modes with either lower or higher wavenumbers and frequencies with respect to the primary modes, providing a mechanism for the bi-directional energy transfer. For instance, at $E=10^{-5}$, $\Pran=0.01$ and $Ra=3.3\times 10^6$, a low-frequency ($\omega/\Omega=0.0918$) prograde mode of $m=3$ excites a pair of retrograde modes with higher frequencies ($\omega/\Omega=0.3596$ and 0.2678) and wavenumbers ($m=10$ and 13). { Note that the primary mode in this case shows a multi-cellular pattern of the radial structure \citep{Net2008}, whereas the primary mode in figure \ref{fig:VelLine} shows a mono-cellular pattern of the radial structure (see supplementary movies).} In this case, we also observe cyclic growth and collapse of the unstable modes (figure \ref{fig:GrowthCollapse}(a-b)), which is a typical behaviour of the triadic resonance \citep{Lin2014,Barker2016}. We can see that higher-wavenumbers modes are excited on top of the $m=3$ primary mode during the growth phase (figure \ref{fig:GrowthCollapse}(c)). As the energy in the growing modes becomes comparable to the zonal flow ($m=0$), the flow suddenly collapses to chaotic small scales (figure \ref{fig:GrowthCollapse}(d)), known as resonant collapse of inertial modes, which was first observed by \cite{McEwan1970} in a precessing cylinder. The collapse is followed by a relaminarization (figure \ref{fig:GrowthCollapse}({ e}), and then the cycle repeats. The whole process is reminiscent of the triadic resonant instability observed in several mechanically driven rotating flows \citep[][and references therein]{LeBars2014}.

Table \ref*{tab:Triads} also lists frequencies of purely inviscid inertial modes in a sphere which have similar spatial structures as the modes observed from numerical simulations. We note that the observed frequencies are slightly shifted from the inviscid frequencies due to the viscous effect, the detuning effect \citep{Kerswell2002} and the coupling with the thermal equation \citep{Zhang1994}.  
Here we report a few cases in which the interacting modes can be clearly identified. For large $Ra$ and low $E$, triadic resonances can simultaneously  excite several pairs of modes, which makes it more difficult to identify individual modes. However, the mechanism appears to be generic as long as the onset of convection is in the form of an oscillatory thermal inertial mode, which generally exists at $Pr<1$ \citep{Zhang1994}.

\begin{figure}
\begin{center}
	\includegraphics[height=5cm]{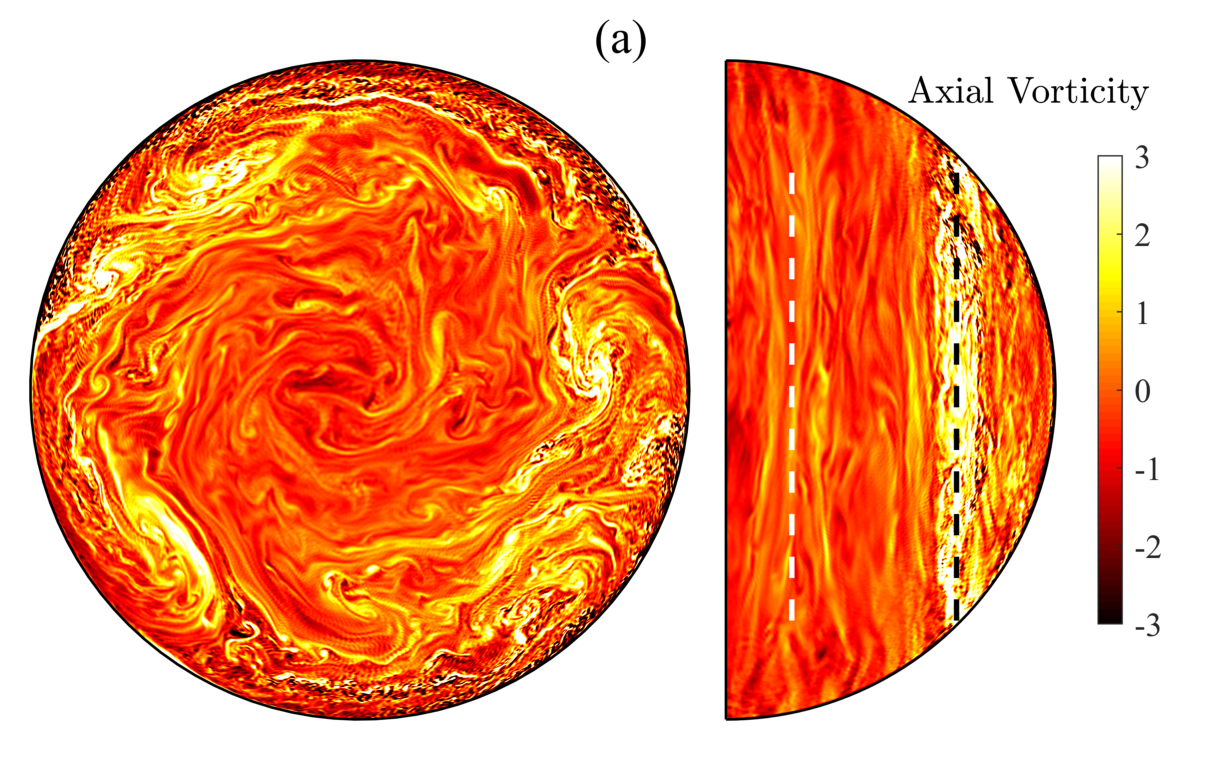}\\
	\includegraphics[height=5.2cm]{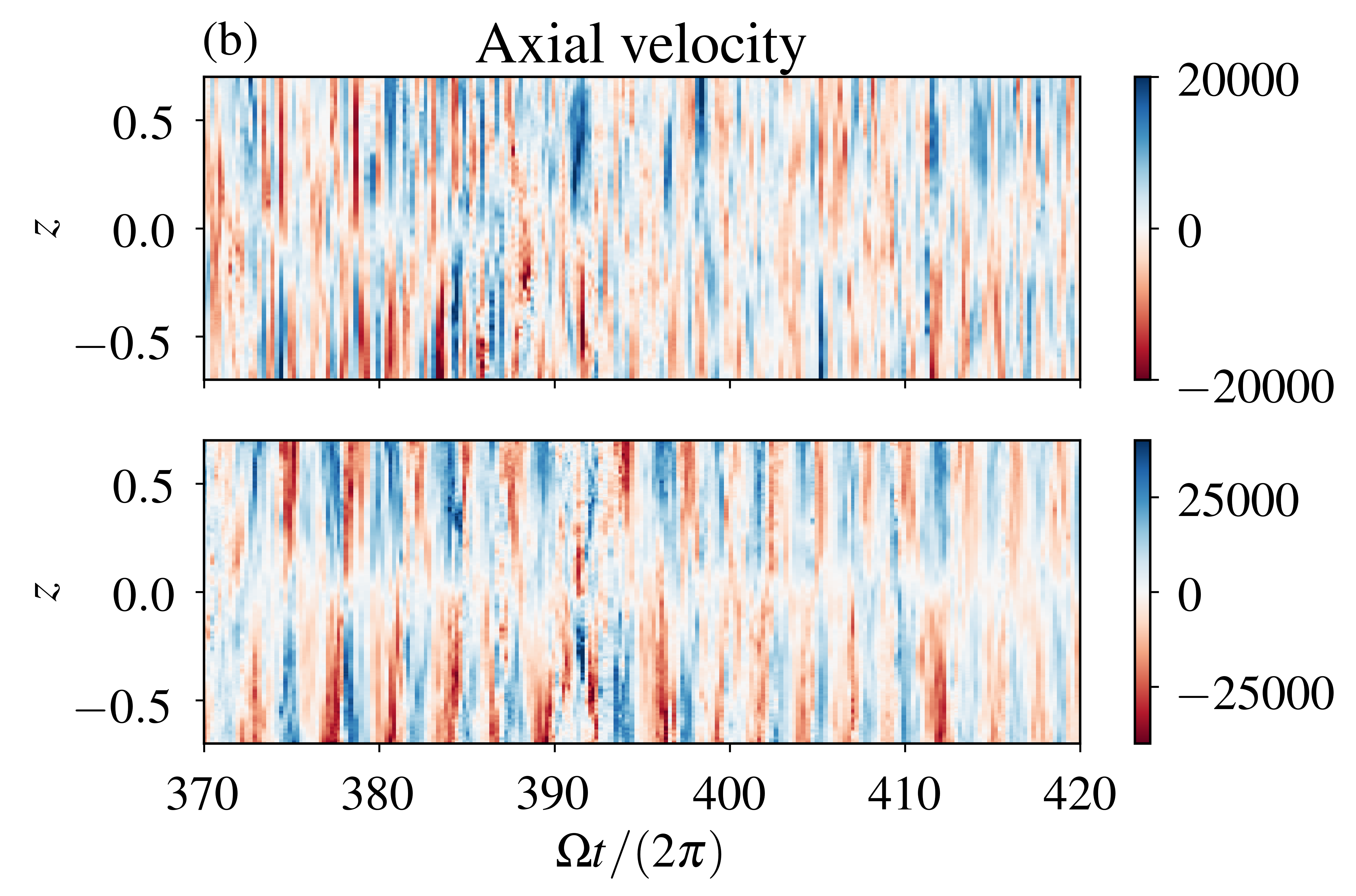} 
	\includegraphics[height=5.2cm]{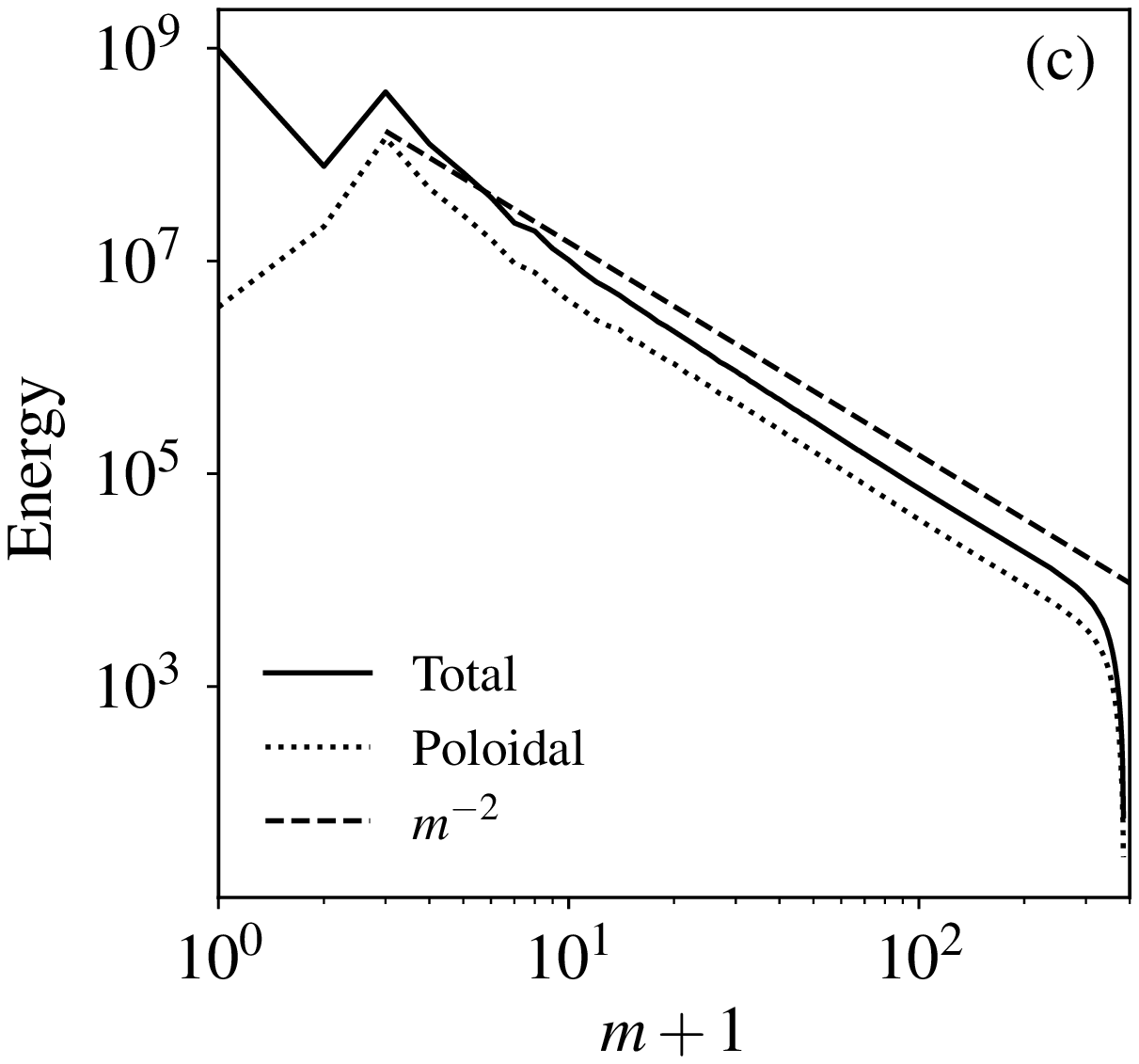}
\end{center}
	\caption{(a) A snapshot of the axial vorticity in the equatorial plane (left) and in a meridional plane (right); (b) Time evolution of the axial velocity along the white (top panel) and black (bottom panel) dashed lines in (a); (c) Energy spectra as a function of the azimuthal wavenubmer $m$. $E=10^{-6}$, $\Pran=0.001$ and $Ra=5.0\times 10^7$.}
	\label{fig:VorZ}
\end{figure}

Further increasing $Ra$ leads to several modes interacting and eventually to a turbulent state. Figure \ref{fig:VorZ} shows an example of turbulent flows at $E=10^{-6}$, $\Pran=0.001$ and $Ra=5.0\times 10^7$. { The Rossby number in this case is calculated to be $Ro=2.9\times10^{-2}$, suggesting that the turbulence convection is strongly influenced by the rotation.} The axial vorticity (figure \ref{fig:VorZ}(a)) exhibits both small-scale and large-scale structures elongated  along the rotation axis. These large-scale vortices located near the equator would merge into a single vortex at the centre when the stress-free boundary condition is used \citep{Lin2020}. Figure \ref{fig:VorZ}(b) shows the time evolution of the axial velocity along two dashed lines in figure \ref{fig:VorZ}(a), which exhibits wave-like behaviours with typical frequencies within the spectrum of inertial waves (see also figure \ref{fig:VelFFT}(c)). Figure \ref{fig:VorZ}(c) shows the energy spectrum as a function of the azimuthal wavenumber $m$. 
We can see from the spectrum that there is a dominant toroidal $m=0$ component, which represents a strong mean zonal flow. 
For smaller scales ($20\lesssim m \lesssim$ 200), we see that the energy spectrum follows a power law of $m^{-2}$, which is reminiscent of the energy spectrum of inertial wave turbulence driven by elliptical instability \citep{LeReun2017}. The power law  $m^{-2}$ is also predicted by \cite{Zhou1995} using a phenomenological approach for rotating turbulence. These observations suggest that the interaction of inertial waves plays an important role even in the turbulent regime, though further investigations are required in future.

\section{Conclusion}
We have shown for the first time that { triadic resonances are taking place} in a rotating convection system. The convective modes can excite either a pair of inertial waves with lower frequencies and wavenumbers or a pair with higher frequencies and wavenumbers, depending on the structure of the primary mode. This provides a { possible} mechanism for { bi-directional} energy transfer. The turbulent regime also shows signatures of inertial waves, and is reminiscent of inertial wave turbulence as observed in mechanically driven rotating flows. This study paves the way to study the nonlinear dynamics and turbulence in rotating convection within the framework of inertial wave interactions. 

{ This study suggests that thermally (at low $Pr$) and mechanically (e.g. tides, precession and libration) driven rotating flows share many features even in the nonlinear and turbulent regime, which may have significant implications for some fundamental problems in geophysical and astrophysical fluid dynamics.  
 Both convection \citep{Jones2015} and mechanical forcings \citep{LeBars2014} are thought to be possible driven mechanisms for the generation of the Earth's and planetary magnetic fields through the dynamo process.  It is possible that dynamos driven by two different mechanisms also share many features in certain parameter regimes and can be studied within the same framework. 
In stars and giant planets, the interaction between convection and tides is of great importance for calculating the efficiency of tidal dissipation, yet the problem remains poorly understood \citep{Ogilvie2014}.  Turbulent convection is usually treated as an effective viscosity in damping the tidal flow \citep{Zahn1989,Vidal2020}. However, given small $Pr$ in the convective zone of  rotating stars and giant planets, both convection and tides can drive inertial waves, which would be interacting simultaneously and make it more difficult to estimate the efficiency of tidal dissipation in these systems. 

In this paper, we consider rotating convection in a  whole sphere. It is also of great interests to consider the problem in a spherical shell, in which behaviours of inertial waves are complicated by the critical latitudes and wave attractors \citep{Rieutord2001}. Therefore, interactions of thermal inertial waves in a spherical shell may exhibit more complex dynamics, which will be considered in future.}

\section*{Declaration of Interests} 
The authors report no conflict of interest.

\section*{Supplementary movies}
Supplementary movies are available at \href{http://faculty.sustech.edu.cn/wp-content/uploads/2020/11/2020111814352427.mp4}{Movie 1}, \href{http://faculty.sustech.edu.cn/wp-content/uploads/2020/11/2020111814352598.mp4}{Movie 2}, \href{http://faculty.sustech.edu.cn/wp-content/uploads/2020/11/2020111814352655.mp4}{Movie 3}.

\section*{Acknowledgements}
The author would like to thank Philippe Marti for providing the numerical code, Andy Jackson for comments on an early draft of this paper, Adrian Barker for useful discussions and the anonymous referees for their helpful comments.
This study was supported by the B-type Strategic Priority Program of the CAS (XDB41000000), an NSFC grant (41904066), and the pre-research project on Civil Aerospace Technologies of CNSA (D020308) and the Macau Foundation. Simulations were performed on the Taiyi cluster supported by the Center for Computational Science and Engineering of Southern University of Science and Technology.

\bibliographystyle{jfm}
\bibliography{reference}

\begin{thebibliography}{38}
\expandafter\ifx\csname natexlab\endcsname\relax\def\natexlab#1{#1}\fi
\def\au#1{#1} \def\ed#1{#1} \def\yr#1{#1}\def\at#1{#1}\def\jt#1{\textit{#1}}
  \def\bt#1{#1}\def\bvol#1{\textbf{#1}} \def\vol#1{#1} \def\pg#1{#1}
  \def\publ#1{#1}\def\arxiv#1{#1}\def\org#1{#1}\def\st#1{\textit{#1}}

\bibitem[Aldridge \& Toomre(1969)]{Aldridge1969}
{\sc \au{Aldridge, K.} \& \au{Toomre, A.}} \yr{1969}  \at{{Axisymmetric
  inertial oscillations of a fluid in a rotating spherical container}}.  \jt{J.
  Fluid Mech.}  \bvol{37},  \pg{307--323}.

\bibitem[Aubert {\em et~al.\/}(2001)Aubert, Brito, Nataf, Cardin \&
  Masson]{Aubert2001}
{\sc \au{Aubert, J}, \au{Brito, D}, \au{Nataf, H.-C.}, \au{Cardin, P} \&
  \au{Masson, J.-P.}} \yr{2001}  \at{{Asystematic experimental study of rapidly
  rotating spherical convection in water and liquid gallium}}.  \jt{Phys. Earth
  Planet. Inter.}  \bvol{128},  \pg{51--74}.

\bibitem[Aurnou {\em et~al.\/}(2015)Aurnou, Calkins, Cheng, Julien, King,
  Nieves, Soderlund \& Stellmach]{Aurnou2015}
{\sc \au{Aurnou, J.M.}, \au{Calkins, M.A.}, \au{Cheng, J.S.}, \au{Julien, K.},
  \au{King, E.M.}, \au{Nieves, D.}, \au{Soderlund, K.M.} \& \au{Stellmach, S.}}
  \yr{2015}  \at{{Rotating convective turbulence in Earth and planetary
  cores}}.  \jt{Phys. Earth Planet. Inter.}  \bvol{246},  \pg{52--71}.

\bibitem[Aurnou {\em et~al.\/}(2018)Aurnou, Bertin, Grannan, Horn \&
  Vogt]{Aurnou2018}
{\sc \au{Aurnou, J.~M.}, \au{Bertin, V.}, \au{Grannan, A.~M.}, \au{Horn, S.} \&
  \au{Vogt, T.}} \yr{2018}  \at{{Rotating thermal convection in liquid gallium:
  multi-modal flow, absent steady columns}}.  \jt{J. Fluid Mech.}  \bvol{846},
  \pg{846--876}.

\bibitem[Barker(2016)]{Barker2016}
{\sc \au{Barker, A.~J.}} \yr{2016}  \at{{Non-linear tides in a homogeneous
  rotating planet or star: global simulations of the elliptical instability}}.
  \jt{Mon. Not. R. Astr. Soc.}  \bvol{459}~(1),  \pg{939--956}.

\bibitem[Bretherton(1964)]{Bretherton1964}
{\sc \au{Bretherton, F.~P.}} \yr{1964}  \at{Resonant interactions between
  waves. the case of discrete oscillations}.  \jt{J. Fluid Mech.}
  \bvol{20}~(3),  \pg{457–479}.

\bibitem[{Busse}(1970)]{Busse1970}
{\sc \au{{Busse}, F.~H.}} \yr{1970}  \at{{Thermal instabilities in rapidly
  rotating systems.}}  \jt{J. Fluid Mech.}  \bvol{44},  \pg{441--460}.

\bibitem[Chandrasekhar(1961)]{Chandra1961book}
{\sc \au{Chandrasekhar, S.}} \yr{1961} {\em Hydrodynamic and hydromagnetic
  stability\/}.  \publ{Clarendon Press}.

\bibitem[Christensen(2002)]{Christensen2002}
{\sc \au{Christensen, U.{\~{}}R.}} \yr{2002}  \at{{Zonal flow driven by
  strongly supercritical convection in rotating spherical shells}}.  \jt{J.
  Fluid Mech.}  \bvol{470},  \pg{115--133}.

\bibitem[Gastine {\em et~al.\/}(2016)Gastine, Wicht \& Aubert]{Gastine2016}
{\sc \au{Gastine, T.}, \au{Wicht, J.} \& \au{Aubert, J.}} \yr{2016}
  \at{{Scaling regimes in spherical shell rotating convection}}.  \jt{J. Fluid
  Mech.}  \bvol{808},  \pg{690--732}.

\bibitem[Godeferd \& Moisy(2015)]{Godeferd2015}
{\sc \au{Godeferd, F.~S.} \& \au{Moisy, F.}} \yr{2015}  \at{{Structure and
  dynamics of rotating turbulence: A review of recent experimental and
  numerical results}}.  \jt{Appl. Mech. Rev.}  \bvol{67}~(3),  \pg{1--13}.

\bibitem[Greenspan(1968)]{Greenspan1968book}
{\sc \au{Greenspan, H.~P.}} \yr{1968} {\em {The Theory of Rotating Fluids}\/}.
  \publ{London: Cambridge University Press}.

\bibitem[Guervilly {\em et~al.\/}(2019)Guervilly, Cardin \&
  Schaeffer]{Guervilly2019}
{\sc \au{Guervilly, C.}, \au{Cardin, P.} \& \au{Schaeffer, N.}} \yr{2019}
  \at{{Turbulent convective length scale in planetary cores}}.  \jt{Nature}
  \bvol{570}~(7761),  \pg{368--371}.

\bibitem[Hollerbach \& Kerswell(1995)]{Hollerbach1995}
{\sc \au{Hollerbach, R.} \& \au{Kerswell, RR}} \yr{1995}  \at{{Oscillatory
  internal shear layers in rotating and precessing flows}}.  \jt{J. Fluid
  Mech.}  \bvol{298},  \pg{327--339}.

\bibitem[Horn \& Schmid(2017)]{Horn2017}
{\sc \au{Horn, S.} \& \au{Schmid, P.~J.}} \yr{2017}  \at{{Prograde, retrograde,
  and oscillatory modes in rotating Rayleigh-B{\'{e}}nard convection}}.  \jt{J.
  Fluid Mech.}  \bvol{831},  \pg{182--211}.

\bibitem[Jones(2015)]{Jones2015}
{\sc \au{Jones, C.A.}} \yr{2015}  \at{{Thermal and Compositional Convection in
  the Outer Core}}.  \bt{In {\em Treatise on Geophysics\/}}, ,  \vol{vol.~8},
  \pg{pp. 115--159}.  \publ{Elsevier}.

\bibitem[Jones {\em et~al.\/}(2000)Jones, Soward \& Mussa]{Jones2000}
{\sc \au{Jones, Chris~A.}, \au{Soward, Andrew~M.} \& \au{Mussa, Ali~I.}}
  \yr{2000}  \at{{The onset of thermal convection in a rapidly rotating
  sphere}}.  \jt{J. Fluid Mech.}  \bvol{405}~(2000),  \pg{157--179}.

\bibitem[Kaplan {\em et~al.\/}(2017)Kaplan, Schaeffer, Vidal \&
  Cardin]{Kaplan2017}
{\sc \au{Kaplan, E.~J.}, \au{Schaeffer, N.}, \au{Vidal, J.} \& \au{Cardin, P.}}
  \yr{2017}  \at{{Subcritical Thermal Convection of Liquid Metals in a Rapidly
  Rotating Sphere}}.  \jt{Phys. Rev. Lett.}  \bvol{119}~(9),  \pg{094501}.

\bibitem[Kerswell(2002)]{Kerswell2002}
{\sc \au{Kerswell, R.~R.}} \yr{2002}  \at{{Elliptical instability}}.  \jt{Annu.
  Rev. Fluid Mech.}  \bvol{34},  \pg{83--113}.

\bibitem[Lam {\em et~al.\/}(2018)Lam, Kong \& Zhang]{Lam2018}
{\sc \au{Lam, Kameng}, \au{Kong, Dali} \& \au{Zhang, Keke}} \yr{2018}
  \at{{Nonlinear thermal inertial waves in rotating fluid spheres}}.
  \jt{Geophys. Astrophys. Fluid Dyn.}  \bvol{112}~(5),  \pg{357--374}.

\bibitem[{Le Bars} {\em et~al.\/}(2015){Le Bars}, C{\'{e}}bron \& {Le
  Gal}]{LeBars2014}
{\sc \au{{Le Bars}, M.}, \au{C{\'{e}}bron, D.} \& \au{{Le Gal}, P.}} \yr{2015}
  \at{{Flows Driven by Libration, Precession, and Tides}}.  \jt{Annu. Rev.
  Fluid Mech.}  \bvol{47}~(1),  \pg{163--193}.

\bibitem[{Le Reun} {\em et~al.\/}(2017){Le Reun}, Favier, Barker \& {Le
  Bars}]{LeReun2017}
{\sc \au{{Le Reun}, T.}, \au{Favier, B.}, \au{Barker, A.~J.} \& \au{{Le Bars},
  M.}} \yr{2017}  \at{{Inertial Wave Turbulence Driven by Elliptical
  Instability}}.  \jt{Phys. Rev. Lett.}  \bvol{119}~(3),  \pg{034502}.

\bibitem[Lin \& Jackson(2021)]{Lin2020}
{\sc \au{Lin, Y.} \& \au{Jackson, A.}} \yr{2021}  \at{{Large-scale vortices and
  zonal flows in spherical rotating convection}}.  \jt{J. Fluid Mech.}
  \bvol{In press}.

\bibitem[Lin {\em et~al.\/}(2014)Lin, Noir \& Jackson]{Lin2014}
{\sc \au{Lin, Y.}, \au{Noir, J.} \& \au{Jackson, A.}} \yr{2014}
  \at{{Experimental study of fluid flows in a precessing cylindrical annulus}}.
   \jt{Phys. Fluids}  \bvol{26}~(4),  \pg{046604}.

\bibitem[Marti \& Jackson(2016)]{Marti2016}
{\sc \au{Marti, P.} \& \au{Jackson, A.}} \yr{2016}  \at{{A fully spectral
  methodology for magnetohydrodynamic calculations in a whole sphere}}.  \jt{J.
  Comput. Phys.}  \bvol{305},  \pg{403--422}.

\bibitem[McEwan(1970)]{McEwan1970}
{\sc \au{McEwan, A.~D.}} \yr{1970}  \at{{Inertial oscillations in a rotating
  fluid cylinder}}.  \jt{J. Fluid Mech.}  \bvol{40}~(03),  \pg{603}.

\bibitem[Net {\em et~al.\/}(2008)Net, Garcia \& S{\'{a}}nchez]{Net2008}
{\sc \au{Net, Marta}, \au{Garcia, Ferran} \& \au{S{\'{a}}nchez, Juan}}
  \yr{2008}  \at{{On the onset of low-Prandtl-number convection in rotating
  spherical shells: Non-slip boundary conditions}}.  \jt{J. Fluid Mech.}
  \bvol{601},  \pg{317--337}.

\bibitem[Noir {\em et~al.\/}(2001)Noir, Brito, Aldridge \& Cardin]{Noir2001b}
{\sc \au{Noir, J.}, \au{Brito, D.}, \au{Aldridge, K.} \& \au{Cardin, P.}}
  \yr{2001}  \at{{Experimental evidence of inertial waves in a precessing
  spheroidal cavity}}.  \jt{Geophys. Res. Lett.}  \bvol{28}~(19),
  \pg{3785--3788}.

\bibitem[Ogilvie(2014)]{Ogilvie2014}
{\sc \au{Ogilvie, G.~I.}} \yr{2014}  \at{{Tidal Dissipation in Stars and Giant
  Planets}}.  \jt{Annu. Rev. Astron. Astrophys.}  \bvol{52},  \pg{1--46}.

\bibitem[Ogilvie \& Lin(2004)]{Ogilvie2004}
{\sc \au{Ogilvie, G.~I.} \& \au{Lin, D. N.~C.}} \yr{2004}  \at{{Tidal
  Dissipation in Rotating Giant Planets}}.  \jt{Astrophys. J.}  \bvol{610}~(1),
   \pg{477--509}.

\bibitem[Plumley \& Julien(2019)]{Plumley2019}
{\sc \au{Plumley, M.} \& \au{Julien, K.}} \yr{2019}  \at{{Scaling Laws in
  Rayleigh‐B{\'{e}}nard Convection}}.  \jt{Earth and Space Science}
  \bvol{6}~(9),  \pg{1580--1592}.

\bibitem[Rieutord {\em et~al.\/}(2001)Rieutord, Georgeot \&
  Valdettaro]{Rieutord2001}
{\sc \au{Rieutord, M.}, \au{Georgeot, B.} \& \au{Valdettaro, L.}} \yr{2001}
  \at{{Inertial waves in a rotating spherical shell: attractors and asymptotic
  spectrum}}.  \jt{J. Fluid Mech.}  \bvol{435},  \pg{103--144}.

\bibitem[Spiegel(1971)]{Spiegel1971}
{\sc \au{Spiegel, E~A}} \yr{1971}  \at{{Convection in Stars I. Basic Boussinesq
  Convection}}.  \jt{Annu. Rev. Astron. Astrophys.}  \bvol{9}~(323).

\bibitem[{Vidal} \& {Barker}(2020)]{Vidal2020}
{\sc \au{{Vidal}, J.} \& \au{{Barker}, A.~J.}} \yr{2020}  \at{{Efficiency of
  tidal dissipation in slowly rotating fully convective stars or planets}}.
  \jt{Mon. Not. R. Astr. Soc.}  \bvol{497}~(4),  \pg{4472--4485}.

\bibitem[{Zahn}(1989)]{Zahn1989}
{\sc \au{{Zahn}, J.~P.}} \yr{1989}  \at{{Tidal evolution of close binary stars.
  I - Revisiting the theory of the equilibrium tide}}.  \jt{Astr. Astrophys.}
  \bvol{220}~(1-2),  \pg{112--116}.

\bibitem[Zhang(1994)]{Zhang1994}
{\sc \au{Zhang, K.}} \yr{1994}  \at{{On coupling between the Poincar{\'{e}}
  equation and the heat equation}}.  \jt{J. Fluid Mech.}  \bvol{268},
  \pg{211}.

\bibitem[Zhang \& Liao(2017)]{Zhang2017}
{\sc \au{Zhang, K.} \& \au{Liao, X.}} \yr{2017} {\em {Theory and Modeling of
  Rotating Fluids: convection, inertial waves, and precession}\/}.
  \publ{Cambridge: Cambridge University Press}.

\bibitem[Zhou(1995)]{Zhou1995}
{\sc \au{Zhou, Y.}} \yr{1995}  \at{{A phenomenological treatment of rotating
  turbulence}}.  \jt{Phys. Fluids}  \bvol{7}~(8),  \pg{2092--2094}.

\end{thebibliography}


\end{document}